\documentclass{article}

\usepackage{amsmath}
\usepackage{amssymb}
\usepackage{amsthm}
\usepackage{array}
\usepackage{enumerate}

\usepackage{microtype}
\usepackage{pgf,tikz}
\usetikzlibrary{shapes,arrows,automata,positioning,backgrounds,fit,decorations.pathreplacing}

\newcommand{\ket}{\rangle}

\newcommand{\coll}[1]{\ensuremath{\left\{ {#1}\right\} }}

\newcommand{\paren}[1]{\ensuremath{\left( {#1} \right)}}
\newcommand{\brac}[1]{\ensuremath{\left[ {#1} \right]}}

\newcommand{\set}[2]{\ensuremath{\left\{\left.#1\,\,\vphantom{#2}\right|\,#2\right\}}}

\newcommand{\mc}[1]{{\mathcal{#1}}}

\newcommand{\Ket}[1]{\left|{#1}\right\ket}
\DeclareMathOperator{\cnot}{CNot}
\newcommand{\triplet}[3]{\brac{{#1},\, {#2},\, {#3}}}
\newcommand{\blank}{\,\rule{5pt}{.5pt}\,}

\newtheorem{claim}{Claim}

\newtheorem{expe}{Experiment}
\newtheorem{theorem}{Theorem}

\begin{document}

\title{Dynamics and Hidden Variables}
\author{Olivier Brunet \\ \texttt{olivier.brunet at normalesup.org}}

\maketitle

\paragraph{Abstract} We study the way the unitary evolution of spin 1/2 particules can be represented in a counterfactual definiteness setting. More precisely, by representing the state of such a particule by a triplet of values corresponding to the supposedly pre-existing outcomes of some measurements (those corresponding to the three Pauli matrices), we analyse the evolution of our representation when some unitary gates (namely, the Hadamard gate, the $\pi$/2 phase shifter and the controlled-not) are applied. Then, we describe in terms of triplets the creation of an EPR pair and discuss the possibility of having this representation comply with the predictions of quantum mechanics. Finally, we show that this is not possible unless one of the assumptions used to build our model is dropped.

\section{Introduction} 
\label{sec:introduction}

Starting with the seminal article of Einstein, Podolsky and Rosen \cite{Einstein35EPR}, many efforts have been done in attempting to complement the orthodox formulation of quantum mechanics with some supplemental data, called \emph{hidden variables}, in order to provide a mechanism that could explain the way measurement outcomes are obtained without renouncing to fundamental physical assumptions such as those of locality or determinism.


A first central result in this domain is provided by Bell's inequalities \cite{Bell64} (reprinted in  \cite{Bell85Book}) which states that one cannot obtain any local hidden variables theory complying with quantum mechanics prediction. This in\-he\-r\-ently probabilistic result has giving birth to several variants, including non-probabilistic ones \cite {CHSH69, GHZ90}.

Another important impossibility result, the Kochen-Specker theorem \cite{Bell66, KochenSpecker67, Mermin93, Brunet07PLA} is usually interpreted as forbidding non-contextual measurements.

All of these results rely on a form of realism, called \emph{counterfactual definiteness} (or \textbf{CFD}), which, quoting \cite{Groblacher2007} (emphasis by the author), ``requires that \textbf{an individual binary measurement outcome} $A$ for a polarization measurement along direction $\vec a$ (that is, whether a single photon is transmitted or absorbed by a polarizer set at a specific angle) \textbf{is predetermined by some set of hidden-variables} $\lambda$, and a three-dimensional vector $\vec u$, as well as by some set of other possibly non-local parameters $\mu$ (for example, measurement settings in space-like separated regions) - that is, $A = A(\lambda, \vec u, \vec a, \mu)$.''

\medskip

In this article, we study the possibility of a hidden-variable model for representing and manipulating spin $\frac 1 2$ quantum particles  (which will hereafter be called qubits) from the dynamics point of view. Our representation will be based on the use of three components, representing the outcomes (thus assumed to exist beforehand) of observables corresponding to the three Pauli matrices. We explore the way some quantum gates, namely the Hadamard gate, the $\frac \pi 2$-phase shifter and the controlled-not gate, can be expressed in this formalism. Then, we use the obtained representations to formalize the creation of a Bell state. Finally, we specify a set of assumptions which were used to justify our constructions, and show that the representation we have provided cannot comply with quantum mechanics predictions, unless one of these assumption is dropped.


\section{Representing States with Triplets} 
\label{sec:triplets_for_representing_qubits_states}

%
%
%
%

Given a qubit, we consider the observables defined by the three Pauli matrices:
$$ \sigma_x = \begin{pmatrix}
	0 & 1 \\ 1 & 0
\end{pmatrix},  \quad \sigma_y = \begin{pmatrix}
	0 & -i \\ i & 0
\end{pmatrix} \quad \hbox{and} \quad \sigma_z = \begin{pmatrix}
	1 & 0 \\ 0 & -1
\end{pmatrix}. $$
Each of those observables has eigenvalues $+1$ and $-1$, with eigenvectors respectively:

$$ \begin{array}{r@{\,=\,}l@{\qquad}r@{\,=\,}l@{\qquad}l}
|X+\ket & \frac 1 {\sqrt 2} \begin{pmatrix}
	1 \\ 1
\end{pmatrix} &
|X-\ket & \frac 1 {\sqrt 2} \begin{pmatrix}
		1 \\ -1
\end{pmatrix} & \hbox{for $\sigma_x$,}\\[10pt]
|Y+\ket & \frac 1 {\sqrt 2} \begin{pmatrix}
	1 \\ i
\end{pmatrix} &
|Y-\ket & \frac 1 {\sqrt 2} \begin{pmatrix}
		1 \\ -i
\end{pmatrix} & \hbox{for $\sigma_y$, and} \\[10pt]
|Z+\ket & \begin{pmatrix}
	1 \\ 0
\end{pmatrix} &
|Z-\ket & \begin{pmatrix}
	0 \\ 1
\end{pmatrix} & \hbox{for $\sigma_z$.} \\
\end{array}
$$

\ 

Following the \emph{counterfactual definiteness} assumption, we will assume that any measurement outcome is precisely defined by some pre-existing property of the particle, independent of the measurement. Thus, given a qubit $Q$, there exists three numbers $x$, $y$ and $z$ belonging to $\coll{-1, +1}$ such that measuring $Q$ with observable $\sigma_x$ (resp. $\sigma_y$, $\sigma_z$) would yield outcome $x$ (resp. $y$, $z$). Using this assumption, we will represent the state of any qubit as a triplet of such values of the form $\triplet x y z$.

\ 

For instance, a qubit in state $|Y-\ket$ would yield $-1$ if measured with $\sigma_y$ and thus will be represented by a triplet of the form $\triplet x {-1} z$ where $x$ and $z$ left as variables, representing definite yet possibly unknown values. Similarly, a qubit in state $|Z+\ket$ will be represented by a triplet of the form $\triplet x y {+1}$.

On occasions, we might write underscores ``\blank'' for some components if they play no role in a given situation. Of course, this does not mean that those values are undefined. For instance, write $\triplet x \blank \blank$ means that we only deal with the value correspond to observable $\sigma_x$.

\ 

Before going any further, we remark that our notation does not make any explicit reference to parameters upon which measurement outcomes may depend. Instead, we only discuss the mere possibility of having a counterfactual definiteness-based theory complying with quantum mechanics predictions: we do not focus on the way these hidden variables are embodied, but rather on the way they behave dynamically.

%


\section{Representing Operators} 

%
%


We now turn to the study of the way three common quantum gates can be represented in terms of triplets. Since triplets represent outcomes of observables $\sigma_x$, $\sigma_y$ and $\sigma_z$, this means that we have to focus on the images of eigenvectors of those observables. By defining
$$\mc B = \coll{|X+\ket, |X-\ket, |Y+\ket, |Y-\ket, |Z+\ket, |Z-\ket},$$
we shall then focus on elements of $\mc B$ (or tensor products of elements of $\mc B$) which are mapped to elements of $\mc B$ up to a phase factor (or, again, to tensor products of such elements).

\subsection{Hadamard} 
\label{sub:hadamard}

Let us first study the Hadamard operator, defined as:
$$ H = \begin{pmatrix}
	1 & 1 \\ 1 & -1
\end{pmatrix} $$
Considering elements of $\mc B$, we have, up to a phase factor:
\begin{gather*}
	H \Ket{X+} = \Ket{Z+} \qquad \qquad H \Ket{X-} = \Ket{Z-} \\
	H \Ket{Y+} = \Ket{Y-} \qquad \qquad H \Ket{Y-} = \Ket{Y+} \\
	H \Ket{Z+} = \Ket{X+} \qquad \qquad H \Ket{Z-} = \Ket{X-}
\end{gather*}
Those equalities can be translated as mapping some values of a representation triplet to others. For instance, the first equality can be expressed as:
$$ H: \triplet {+1} \blank \blank \mapsto \triplet \blank \blank {+1} $$
We can express all the previous equalities in a similar way way:
$$ H: \left\{ \begin{array}{r@{\ \mapsto\ }l}
	\triplet {+1} \blank \blank & \triplet \blank \blank {+1} \\
	\triplet {-1} \blank \blank & \triplet \blank \blank {-1} \\
	\triplet \blank {+1} \blank & \triplet \blank {-1} \blank \\
	\triplet \blank {-1} \blank & \triplet \blank {+1} \blank \\
	\triplet \blank \blank {+1} & \triplet {+1} \blank \blank \\
	\triplet \blank \blank {-1} & \triplet {-1} \blank \blank \\
\end{array} \right. $$

Let us now study the way these mappings lead to a function-like representation of the Hadamard operator acting on triplets. Consider, for instance, a triplet $\triplet x y z$ representing a qubit and let $\triplet {x'} {y'} {z'}$ represent the same qubit, but after having been applied the Hadamard operator. Equality $H |X+\ket = |Z+\ket $ translates~as:
$$ x = +1 \Rightarrow z' = +1. $$
Similarly, equality $H |X-\ket = | Z- \ket $ leads to:
$$ x = -1 \Rightarrow z' = -1. $$
But considering the contraposition of the first implication, one gets: 
$$z' \neq +1 \Rightarrow x \neq +1.$$
Since both $x$ and $z'$ belong to $\coll{-1,+1}$, this is equivalent to:
$$z' = -1 \ \Rightarrow\  x = -1.$$
This means that we actually have an equivalence:
$$ x = -1 \iff z' = -1. $$
Considering the first implication and the contraposition of the second, we also deduce that $x = +1 \iff z' = +1$. In other words, one has $z' = x$. Similarly, one has $y' = -y$ and $x' = z$. Thus, it is possible to represent the action of the Hadamard operator in terms of triplets as the following function:
$$ h: \triplet x y z \mapsto \triplet z {-y} x $$
Such a function will be called a \emph{faithful functional representation} of the Hada\-mard gate. Here, \emph{faithful} means that its definition is entirely based on the direct translation of equalities such as $H |Z+\ket = |X+\ket$ in terms on triplets. The adjective \emph{functional} means that we have obtained sufficiently many relations so that the resulting triplet does entirely depend on the argument triplet, i.e.\ there is a funct\-ional type of relation between the argument and the result triplets.





\subsection{Phase Shifter} 
\label{sub:phase_shifter_}

Let us now consider the $\frac \pi 2$-phase shifter:
$$ P_{\frac \pi 2} = \begin{pmatrix}
	1 & 0 \\ 0 & i
\end{pmatrix} $$
Up to phase factors, one has:
\begin{gather*}
	P_{\frac \pi 2} \Ket{X+} = \Ket{Y+} \quad P_{\frac \pi 2} \Ket{X-} = \Ket{Y-} \\
	P_{\frac \pi 2} \Ket{Y+} = \Ket{X-} \quad P_{\frac \pi 2} \Ket{Y-} = \Ket{X+} \\
	P_{\frac \pi 2} \Ket{Z+} = \Ket{Z+} \quad P_{\frac \pi 2} \Ket{Z-} = \Ket{Z-}
\end{gather*}
By proceeding as before, one finds that the $\frac \pi 2$-phase shifter has the following faithful functional representation:
$$ p_{\frac \pi 2}: \triplet x y z \mapsto \triplet {-y} x z $$

\subsection{Controlled-Not} 
\label{sub:controlled_not}

Finally, we focus on the representation of the action of the controlled-not operator in terms of triplets. This operator is defined by the following matrix:
$$ \cnot = \begin{pmatrix}
	1 & 0 & 0 & 0 \\ 0 & 1 & 0 & 0 \\ 0 & 0 & 0 & 1 \\ 0 & 0 & 1 & 0 \\
\end{pmatrix}. $$
Contrary to the previous operators, a tensor product of elements of $\mc B$ might not be mapped to such a tensor product. For instance:
$$ |Y+,Y+\ket = \frac 1 2 \begin{pmatrix}
	1 \\ i \\ i \\ -1 
\end{pmatrix} \hbox{\quad is mapped to \quad} \cnot |Y+,Y+\ket = \frac 1 2 \begin{pmatrix}
		1 \\ i \\ -1 \\ i 
	\end{pmatrix} $$
The resulting vector cannot be expressed as a tensor product of two two-dimensional vectors, and hence does not belong to $\mc B^{\otimes 2} = \set {u \otimes v}{(u,v) \in \mc B^2}$.

\medskip

An exhaustive list of mapping from $\mc B^{\otimes 2}$ to itself (again, up to a phase factor)~is:

$$ \begin{array}{c@{\,}l@{\,}l@{\quad}c@{\,}l@{\,}l}
\cnot & |Z+,Y+\ket & = |Z+,Y+\ket & \cnot & |Z+,Y-\ket & = |Z+,Y-\ket \\
\cnot & |Z+,Z+\ket & = |Z+,Z+\ket & \cnot & |Z+,Z-\ket & = |Z+,Z-\ket \\
\cnot & |Z-,Y+\ket & = |Z-,Y-\ket & \cnot & |Z-,Y-\ket & = |Z-,Y+\ket \\
\cnot & |Z-,Z+\ket & = |Z-,Z-\ket & \cnot & |Z-,Z-\ket & = |Z-,Z+\ket \\[6pt]
\cnot & |Z+,X+\ket & = |Z+,X+\ket & \cnot & |Z+,X-\ket & = |Z+,X-\ket \\
\cnot & |Z-,X+\ket & = |Z-,X+\ket & \cnot & |Z-,X-\ket & = |Z-,X-\ket \\[6pt]
\cnot & |X+,X+\ket & = |X+,X+\ket & \cnot & |X+,X-\ket & = |X-,X-\ket \\
\cnot & |X-,X+\ket & = |X-,X+\ket & \cnot & |X-,X-\ket & = |X+,X-\ket \\
\cnot & |Y+,X+\ket & = |Y+,X+\ket & \cnot & |Y+,X-\ket & = |Y-,X-\ket \\
\cnot & |Y-,X+\ket & = |Y-,X+\ket & \cnot & |Y-,X-\ket & = |Y+,X-\ket \\
\end{array} $$

Several remarks can be made concerning those equalities:
\begin{enumerate}
	\item The $Z$-component of the first qubit and the $X$-component of the second qubit are left unchanged. Functionally, this can be written~as:
	$$ \mathit{cnot}\,\paren{\triplet \blank \blank {z_1},\triplet {x_2} \blank \blank} \mapsto \paren{\triplet \blank \blank {z_1},\triplet {x_2} \blank \blank}. $$
	\item The $Z$-component of the first qubit determines the change of the $Y$- and $Z$-components of the second qubit:
	$$ \mathit{cnot}\,\paren{\triplet \blank \blank {z_1}, \triplet \blank {y_2} {z_2}} \mapsto \paren{\triplet \blank \blank {z_1}, \triplet \blank {z_1 y_2} {z_1 z_2}}. $$
	\item Similarly, the $X$-component of the second qubit determines the change of the $X$- and $Y$-components of the first one:
	$$ \mathit{cnot}\,\paren{\triplet {x_1} {y_1} \blank, \triplet {x_2} \blank \blank} \mapsto \paren{\triplet {x_1 x_2} {y_1 x_2} \blank, \triplet {x_2} \blank \blank}. $$
\end{enumerate}

By merging these relations, it is clear that $X$-, $Y$- and $Z$-components of each qubit after applying the controlled-not operator are completely determined by the components before the operator was applied. Hence, the controlled-not operator also has a faithful functional representation in terms of triplets: 
$$ \mathit{cnot}: \paren{\triplet {x_1} {y_1} {z_1}, \triplet {x_2} {y_2} {z_2}} \mapsto \paren{\triplet {x_1 x_2} {y_1 x_2} {z_1}, \triplet {x_2} {z_1 y_2} {z_1 z_2}}. $$


\bigskip

With the previous study, we have obtained a faithful functional representation of three operators: the Hadamard gate, the $\frac \pi 2$-phase shifter and the controlled-not. Those operators are sufficient to describe the production an EPR pair.

In the next section, we will study the way the production of such a pair can be represented in terms of triplets. Then, we will show that our model is in contradiction with quantum mechanics predictions and will attempt to provide some interpretational elements about this contradiction.



\section{The Production of an EPR Pair} 

An EPR pair, composed of two maximally entangled qubits, is defined by Bell state $|\Psi^-\ket$:
$$ |\Psi^-\ket = \frac 1 {\sqrt 2} \paren{|01\ket - |10\ket}$$
Such a state can be obtained using the following quantum circuit:

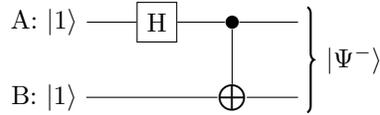
\hbox{} \hfill
\begin{tikzpicture}

\tikzstyle{operator} = [draw,fill=white,minimum size=1.5em] 
\tikzstyle{phase} = [fill,shape=circle,minimum size=5pt,inner sep=0pt]
\tikzstyle{surround} = [fill=blue!10,thick,draw=black,rounded corners=2mm]

\node (q_0) at (-0.5,0) {A:~$|1\ket$} ;
\node (q_4) at (-0.5,-1) {B:~$|1\ket$} ;

\node[operator] (q_1) at (1,0) {H} edge [-] (q_0);
\node[phase] (q_2) at (2,0) {} edge [-] (q_1);
\node (q_3) at (3,0) {} edge [-] (q_2) ;
\node[inner sep = 0pt, outer sep = 0pt] (q_6) at (2,-1) {$\bigoplus$} edge [-] (q_4);
\node (q_7) at (3,-1) {} edge [-] (q_6) ;

\draw (q_2) -- (q_6) ;

\draw[decorate,decoration={brace},thick] (3,0.2) to
	node[midway,right] (bracket) {$\ |\Psi^-\ket$}
	(3,-1.2);
	
\end{tikzpicture} \hfill \hbox{} \\
This circuit involves two qubits, called $A$ and $B$. We will represent their states by a pair of two triplets $(\triplet {x_1} {y_1} {z_1}, \triplet {x_2} {y_2} {z_2})$ where $\triplet {x_1} {y_1} {z_1}$ (resp. $\triplet {x_2} {y_2} {z_2}$) corresponds to $A$ (resp. $B$). The previous circuit translates in terms of triplet the following~way:
\begin{enumerate}
	\item The initial qubits are represented by triplet of the form $\triplet \blank \blank {-1}$, that is $z_1 = z_2 = -1$. The initial system can thus be written~as:
	$$ \paren{\triplet {x_1} {y_1} {-1}, \triplet {x_2} {y_2} {-1}} $$
	with $x_1$, $x_2$, $y_1$ and $y_2$ belonging to $\coll{-1, 1}$.
	\item After applying the Hadamard gate to $A$, one obtains:
	$$ \paren{\triplet{-1} {-y_1} {x_1}, \triplet {x_2} {y_2} {-1}} $$
	\item Finally, the application of the controlled-not leads~to:
$$ \paren{\triplet {-x_2} {-y_1 x_2} {x_1}, \triplet {x_2} {x_1 y_2} {-x_1}}. $$
\end{enumerate}
By construction, the obtained pair of triplets is the general form for representing the $|\Psi^-\ket$ state in terms of triplets.

%


\section{Contradiction} 

Quantum mechanics predicts that if the spin of each qubit of a $|\Psi^-\ket$ Bell state pair is measured along the same direction, opposite outcomes are to be obtained. In terms of triplets, this means that the values of a given component in the previous pair of triplets must be opposite.

It is clearly the case for the $X$-components (where we have $-x_2$ and $x_2$) and the $Z$-components (where we have $x_1$ and $-x_1$).

\ 

Let us now focus on the $Y$-components. The corresponding value of the first qubit is of the form $- y_1 x_2$ while that of the second qubit is $x_1 y_2$. In order to agree with quantum mechanics, one must~have:
$$ y_1 x_2 = x_1 y_2 $$
By multiplying both side of this equality by $x_1 x_2$, since ${x_1}^{\!2} = {x_2}^{\!2} = 1$, one~gets:
$$ x_1 y_1 = x_2 y_2 $$
Each side of this equality describes the same quantity for each qubit. If we call that quantity the XY-product of a qubit, we have:

\begin{claim}
Any pair of $|Z-\ket$-state qubits used to produce a $|\Psi^-\ket$ Bell state must have equal XY-products.
\end{claim}

Now, we turn to the way the $\frac \pi 2$-phase shifter acts on triplets. We recall that it is represented by function $p_{\frac \pi 2}: \triplet x y z  \mapsto \triplet y {-x} z$. Applying it to a qubit in state $|Z-\ket$, one gets:

\begin{claim}
The application of a $\frac \pi 2$-phase shifter to a qubit in state $|Z-\ket$ yields a qubit, also in state $|Z-\ket$, with opposite XY-product.
\end{claim}

The combination of those two claims leads to a contradiction, as illustrated by the following experimental setup:

\begin{expe}
	This experiment is composed of three steps, and is described as performed by Alice:
	\begin{enumerate}[\hspace{0.5cm}Step 1.]
		\item Alice creates two qubits $A$ and $B$, both in state $|Z-\ket$;
		\item Then, she chooses whether she applies a $\frac \pi 2$-phase shifter to qubit $A$;
		\item Finally, she produces a $|\Psi^-\ket$ state with $A$ and $B$ using the previous circuit.
	\end{enumerate}
\end{expe}


\medskip

Suppose that Alice has chosen not to apply the phase shifter. Following Claim 1, at the beginning of Step~3., qubits $A$ and $B$ must have equal XY-products. This means that already at the end of Step~1., both qubits have equal XY-products.

Now, suppose instead that Alice has chosen to apply the phase shifter. Then again, at the beginning of Step~3., qubits $A$ and $B$ must have equal XY-products. But, because of the application of the phase shifter, at the end of Step~1., both qubits must this time have opposite XY-products.

\medskip

But since Step 1.\ happens \emph{before} Alice has chosen whether she would apply the phase shifter, both conditions (whether XY-products are equal or opposite) should hold simultaneously, which is not possible, hence the announced contradiction.



\section{Main Result} 
\label{sec:discussion}

Now that our contradiction has been presented, we study more precisely the assumptions which have led to it. The first one has already been specified:
\begin{description}
	\item[CFD] (Counterfactual Definiteness) Any measurement outcome is precisely defined by some pre-existing property of the particle, independent of the measurement.
\end{description}
This has allowed us to define the triplet representation of a qubit.

\ 

The next step has been to define what we have called \emph{faithful functional representations} of some quantum operators. For instance, considering the Hadamard gate, equalities like $H |X-\ket = |Z-\ket$ or $H |Y+\ket = |Y-\ket$ were translated in terms of triplets:
$$ \triplet {-1} \blank \blank \mapsto \triplet \blank \blank {-1} \qquad \triplet \blank {+1} \blank \mapsto \triplet \blank {-1} \blank $$
Then merging all the obtained relations, we have obtained our functional representation:
$$ h~: \triplet x y z \mapsto \triplet z {-y} x $$
But this ``merging'' operation can be regarded as doubtful, since while a single relation like $\triplet {-1} \blank \blank \mapsto \triplet \blank \blank {-1}$ is the direct translation of a testable and falsifiable quantum fact, we lose this testability if we consider the three triplet values at once. This means that we have made another assumption, which we propose to express~as:
\begin{description}
	\item[MAD] (Measurement-Agnostic Dynamics) The modification of any value of a triplet, when applying an unitary operator, is independent of the fact that this value has been measured or, more generally, that it can be known with absolute certainty without interacting with the particle.
\end{description}
This assumption makes our construction of functional representations valid, because, writing $h(\triplet x y z) = \triplet {x'} {y'} {z'}$, equality $H |X-\ket = |Z-\ket$ translates~as:
$$ x=-1 \ \Rightarrow z'=-1 $$
and that implication is considered as valid in every situation. 

\ 

Finally, during Step 2.\ of the described preparation, where Alice could choose whether a phase shifter is applied, we implicitly used a final assumption, that of \emph{free will}, a notion has gained much attention lately \cite{ConwayKochen06:FreeWill, ConwayKochen09:FreeWill, GoldstainTauskTumulkaZanghi10:FreeWill, tHooft07:FreeWill}. For the present article, what we mean by free will is that it is possible to be in a situation where the choice of applying the phase shifter can be made \emph{independently} of the qubits' states. More precisely, in our presentation of the contradiction, where we have stated that the situation at Step 1.\ happened before Alice had chosen whether she would apply the phase shifter, what should be understood is that both elements should (or, at least, could) be independent from one another. In a setting where locality holds, it is sufficient to have the preparation of the entangled particles on the one hand, and the choice regarding the application of the phase shifter on the other hand be done in two space-like separated regions. Formally, we express this requirement as:
\begin{description}
	\item[FW] (Free-Will) There exists two triplets $T_1 = \triplet {x_1}{y_1}{-1}$ and $T_2 = \triplet {x_2}{y_2}{-1}$ such that it is possible to use two qubits, described by $T_1$ and $T_2$ respectively, in the previous experiment and such that both possibilities regarding the phase shifter (that is, whether it is applied or not) may occur.
\end{description} 

These assumptions suffice to express and justify our definition of triplets and of functional representations, together with the construction of the previous contradiction. We thus have proved the following result.

\begingroup
\begin{theorem}\label{thm:NoGo}
	No physical theory verifying \textbf{CFD}, \textbf{MAD} and \textbf{FW} can reproduce all the predictions of Quantum Mechanics.
\end{theorem}
\endgroup

\section{Discussion}

%

Let us first focus on free will. 
%
In our discussion, this assumption only reflects the fact that the choice of applying the phase-shifter can be made independently of the qubits used to make the EPR pair. More specifically, it assumes that the following two alternatives are independent from one another: on the one hand whether the $XY$-product of the qubits involved in the EPR pair are equal, and on the other hand whether the phase-shifter is applied.

This type of assumption is central in science: we suppose the existence of \emph{a priori} distinct and independent entities in such a way that if two such independent entities interact, correlations between measurable quantities of those two entities should only be a consequence of the interaction.

Suppose otherwise. In that case, with the further assumption that \textbf{CFD} and \textbf{MAD} both hold, it would not be possible to have any independence between the equality of the $XY$-products and the potential application of the phase-shifter. On the contrary, there would always be a correlation so as to make \textbf{Claim 1} apparently true. But such an always-existing correlation would be all the more surprising, considering that quantum mechanics appears to be fundamentally random regarding measurement outcomes (and hence triplet values in our setting).




Thus, if there really existed such connections making us believe that measurement outcomes are actually random while make \textbf{Claim 1} apparently true, this would lead to the necessary abandonment of some important elements of the scientific method: we would have the false impression that there exists independent entities, while being tricked into not being able to falsify this impression. As a consequence, while we cannot prove that our free will assumption actually holds, we have to believe it does if we are to carry on with any scientific activity.

\ 

Now, suppose that \textbf{FW} actually holds (which we consider, from our previous discussion, as a necessity) and let's turn to \textbf{CFD} and \textbf{MAD}. Since our \textbf{MAD} assumption needs counterfactual definiteness as a premise, dropping \textbf{CFD} would imply to drop \textbf{MAD} also.

But what would it mean to drop \textbf{MAD} alone? To answer this, let us first recall that what has led us to consider this assumption was the necessity for the unitary evolution of triplet values (or, more generally, of possibly hidden yet measurable values in a hidden variables context) to be independent of whether a value has been measured or, more generally, is \emph{knowable}. What we mean by this is that a given value can be deduced from earlier measurement outcomes. For instance, if, after a measurement, a qubit is known to be in state $|0\ket$, then its $Z$-value is knowable: it is possible to know with certainty the outcome of a measurement yielding its $Z$-value. Now, if an Hadamard gate is applied to this qubit, its $X$ value becomes knowable while its $Z$ value is now longer so. Similarly, given an EPR-pair in state $|\Psi^-\ket$ made of particles $A$ and $B$, if $A$ is measured, yielding its $Z$-value, then the $Z$-value of particle $B$ becomes knowable. But if the measurement of particle $A$ occurs in a situation where the outcome cannot be transmitted (the laboratory may explode, or be caught in a black hole), would it still be true that the $Z$-value of particle $B$ is knowable?

Thus, if the \textbf{MAD} assumption were to be dropped, one would need to provide a general and rigorous definition of this notion of \emph{knowability}, and the previous examples show that this might not be an easy task. Of course, this follows directly from the fact that the notion of knowability is closely related to that of measurement, and theses examples only illustrate the fact that the latter remains extremely difficult to grasp. Yet, an important motivation for the  hidden variables program is avoid some difficultes inherent to this notion, and the dropping of the \textbf{MAD} assumption would imply the necessity to actually tackle those difficulties rather than ignoring them.

Finally, suppose (still in the situation where the \textbf{MAD} assumption is dropped) that it is possible to devise a satisfactory definition of \emph{knowability} (and thus, probably, of measurement). In that case, the unitary evolution of measurement values would depend on whether they are knowable. But then, in order to built a counterfactual definiteness theory providing strictly more results that standard quantum mechanics -- which would \emph{a priori} amout to providing results about un-knowable values -- one would need to devise the unitary behavior of measurement values even in the case where they are un-knowable. This means that instead of eluding some difficulties inherent to the notion of measurement, the elaboration of a \textbf{CFD} theory would result in tackling those difficulties, together with some additional ones.

\ 

In conclusion, we consider that \textbf{FW} is a necessary assumption for any scientific theory. Now, in order to keep \textbf{CFD}, one should drop the \textbf{MAD} assumption, which would imply the necessity of defining precisely the notion of knowability (which is the core of the \textbf{MAD} assumption). Moreover, even if this were achieved, the obtention of an \emph{interesting}  \textbf{CFD} theory (insofar as it would make strictly more predictions that standard quantum mechanics) remains subordinated to the getting of results about the unitary evolution of unknowable quantities. To that respect, while one cannot prove that the impossibility of obtaining a counterfactual definiteness theory compatible with the predictions of quantum mechanics, 
we strongly believe that, due to the necessary dropping of the \textbf{MAD} assumption, the expected benefits of obtaining such a theory should be greatly lowered.

\bibliographystyle{apalike}

\end{document}